\documentclass[final,5p,times,twocolumn]{elsarticle}
\usepackage[dvipsnames]{xcolor}
\usepackage{bm}
\usepackage{graphicx}
\usepackage{amssymb}
\usepackage{amsmath}
\usepackage{hyperref}
\usepackage{caption}
\usepackage{float}
\usepackage{slashed}
\usepackage{comment}
\usepackage{cuted}

\makeatletter 
\def\ps@pprintTitle{%
 \let\@oddhead\@empty 
\let\@evenhead\@empty 
\let\@oddfoot\@empty 
\let\@evenfoot\@oddfoot }
\makeatother

\newcommand{\eps}{\varepsilon}

\newcommand{\invM}{\hspace{-.05cm}\int \hspace{-.1cm}\frac{d N}{2\pi i} x^{-N}}
\newcommand{\mell}[1]{#1}
\newcommand{\Kdis}{K_{\mathchoice
  {\mathrm{DIS}}
  {\mathrm{DIS}}
  {\scriptscriptstyle\mathrm{DIS}}
  {\scriptscriptstyle\mathrm{DIS}}}}
\newcommand{\Ktmd}{K_{\mathchoice
  {\mathrm{TMD}}
  {\mathrm{TMD}}
  {\scriptscriptstyle\mathrm{TMD}}
  {\scriptscriptstyle\mathrm{TMD}}}}
\newcommand{\Kgen}{K_{\text{\footnotesize gen}}}

\journal{Physics Letters B}

\begin{document}

\begin{frontmatter}

\title{Unveiling the Collins-Soper kernel in inclusive deep inelastic scattering at threshold}

\author[a]{Andrea Simonelli\fnref{orcid_a}} 
\fntext[orcid_a]{ORCID: 0000-0003-2607-9004} \ead{andrea.simonelli@roma1.infn.it}
\author[b1,b2]{Alberto Accardi\fnref{orcid_b}} 
\fntext[orcid_b]{ORCID: 0000-0002-2077-6557} 
\ead{accardi@jlab.org} 
\author[b1,b2]{Matteo Cerutti\fnref{orcid_c}} 
\fntext[orcid_c]{ORCID: 0000-0001-7238-5657} 
\ead{matteo.cerutti@cea.fr} 
\author[d1,d2]{Caroline S. R. Costa\fnref{orcid_d}} 
\fntext[orcid_d]{ORCID: 0000-0003-1392-837X} 
\ead{costa@berkeley.edu} 
\author[e1,e2]{Andrea Signori\fnref{orcid_e}} 
\fntext[orcid_e]{ORCID: 0000-0001-6640-9659} 
\ead{andrea.signori@unito.it} 

\affiliation[a]{organization={INFN Sezione di Roma}, addressline={Piazzale Aldo Moro 5}, city={Roma}, postcode={00185}, country={Italy} }

\affiliation[b1]{ organization={Christopher Newport University}, addressline={}, city={Newport News}, postcode={23606}, state={Virginia}, country={USA} }
\affiliation[b2]{ organization={Jefferson Lab}, addressline={}, city={Newport News}, postcode={23606}, state={Virginia}, country={USA} }

\affiliation[d1]{ organization={Department of Physics, University of California}, city={Berkeley}, postcode={94720}, state={California}, country={USA} }

\affiliation[d2]{ organization={Nuclear Science Division, Lawrence Berkeley National Laboratory}, city={Berkeley}, postcode={94720}, state={California}, country={USA} }

\affiliation[e1]{organization={Department of Physics, University of Turin}, addressline={via Pietro Giuria 1}, city={Turin}, postcode={10125}, country={Italy} }

\affiliation[e2]{organization={INFN Sezione di Torino}, addressline={via Pietro Giuria 1}, city={Turin}, postcode={10125}, country={Italy} \\[-0.3cm] }

\begin{abstract}
We revisit the factorization of inclusive deep inelastic scattering (DIS) near the kinematic threshold in terms of collinear, off-light-cone operators. 
At threshold, particle production develops around two opposite near-light-cone directions in close analogy with transverse-momentum-dependent semi-inclusive DIS. 
The Collins-Soper kernel then emerges as the universal function governing the rapidity evolution of the relevant parton correlators in both cases. 
Our new framework also clarifies outstanding issues related to soft radiation and rapidity divergences at threshold.
\end{abstract}

\begin{keyword}
Deep inelastic scattering \sep Threshold factorization \sep Collins-Soper kernel \sep Rapidity evolution 
\\
\textit{Preprint:} JLAB-THY-26-4859
\end{keyword}

\end{frontmatter}

\section{Introduction}
Factorization of the electron-proton deep inelastic scattering (DIS) cross section is a testing ground for Quantum Chromodynamics (QCD) and is essential for separating short-distance, perturbatively calculable electron-quark interactions from the long-distance structure of the proton \cite{Pennington:1982kr,Collins:2011zzd,Blumlein:2012bf}. 
The latter is encoded in nonperturbative parton distribution functions (PDFs), whose accurate determination is crucial for describing high-energy collisions at the Large Hadron Collider and in the search for physics beyond the Standard Model \cite{Kovarik:2019xvh,Narain:2022qud,Salam:2022izo,Bacchetta:2023hlw,Ubiali:2024pyg}. 
Improving this accuracy is particularly critical near the kinematic threshold, where the partons fractional momentum relative to the proton is large. 
In this regime, PDFs are poorly constrained by current data and rank as one of the biggest source of uncertainties in various QCD calculations~\cite{Ball:2022qtp,Gao:2017yyd, Beenakker:2015rna,Fiaschi:2022wgl,Brady:2011hb}. 
The large parton momentum region also provides a unique window on color confinement and other nonperturbative phenomena such as quark-hadron duality \cite{Bloom:1970xb,Bloom:1971ye,Melnitchouk:2005zr,Holt:2010vj}, and motivates experimental and phenomenological efforts at present and future facilities~\cite{Dudek:2012vr,Hadjidakis:2018ifr,Accardi:2023chb}.

Factorization of inclusive DIS in the threshold region has been extensively discussed within the framework of soft-collinear effective field theories (SCET)~\cite{Manohar:2003vb, Manohar:2005az,Pecjak:2005uh,Chay:2005rz,Becher:2006nr,Chen:2006vd,Becher:2006mr,Chay:2012jr,Chay:2013zya,Chay:2017bmy} shedding new light on earlier work in QCD~\cite{Sterman:1986aj, Catani:1989ne, Korchemsky:1992xv}. Yet, questions remain open regarding the universality of the soft function and the delicate treatment of rapidity divergences~\cite{Fleming:2012kb,Fleming:2016nhs}.
In this Letter we propose an off-light-cone approach to collinear factorization that explicitly addresses and resolves these issues by exploiting geometrical similarities between the kinematic configurations that dominate DIS at threshold and those dominating the transverse-momentum-dependent (TMD) cross section in semi-inclusive DIS (SIDIS). 
The theorem we obtain encompasses previous results and unveils how the Collins-Soper kernel~\cite{Collins:2011zzd,Collins:1981uk,Collins:1981va,Collins:1984kg,Vladimirov:2020umg,BermudezMartinez:2022ctj,Moos:2023yfa,Bacchetta:2024qre} also describes soft gluon interactions in inclusive processes at threshold. 

\section{TMD factorization features in inclusive DIS}
As the Bjorken variable approaches its kinematic limit, ${x \to 1}$, the invariant mass $W$ of the process decreases, severely restricting particle production in the final state. 
In this regime, radiation cannot form any collimated direction except those associated with the fragmentation of the struck quark and with the target remnant. 
In the Breit frame, these two directions are opposite to each other and approach the light cone when masses are negligible compared to the hard scale $Q$.
The same configuration also controls the transverse-momentum distribution of the observed hadron in SIDIS~\cite{Collins:1989gx,Bacchetta:2006tn,Bacchetta:2008xw, Becher:2010tm,Collins:2011zzd,Aybat:2011zv, Echevarria:2011epo}, suggesting that the two processes share analogous factorization properties. 
We will exploit this common geometrical structure to derive a TMD-like factorization theorem for the unpolarized DIS hadronic tensor at threshold:
\begin{align}
    \label{eq:fact_theo_sqrtdef}
    W^{\mu\nu} = 
    \mathbf{u}^{\mu \nu} 
    H_j(\mu,Q) 
    \invM
    \mell{\bm{\mathcal{F}}}_{j}(N;\mu,y_n) 
    \mell{\bm{\mathcal{J}}}_j(N;\mu,y_n) \ ,
\end{align}
where repeated indices imply, as throughout this paper, a sum over repeated quark flavor indexes $j$ weighted by their squared charges, $N$ is the Mellin space variable, and we have defined the tensor ${\mathbf{u}^{\mu \nu}  = \delta^{\mu +} \delta^{\nu -}+\delta^{\mu -} \delta^{\nu +} -g^{\mu \nu}}$~\footnote{The $\mathbf{u}^{\mu \nu}$ tensor may be replaced by $\mathbf{p}^{\mu \nu}  =- i \epsilon^{+ - \mu \nu}$ for polarized scattering.}. 
$H_j$ is a perturbatively calculable function describing the interaction of the virtual photon with a target quark, $\mu$ is the renormalization scale, and $y_n$ is a rapidity scale that characterizes the nonperturbative, \textit{off-light-cone} target function $\mell{\bm{\mathcal{F}}}_{j}$ and \textit{off-light-cone} inclusive jet function $\mell{\bm{\mathcal{J}}}_j$. 
These functions play a role analogous to that of TMD parton distribution and fragmentation functions in SIDIS, and we will show that they likewise satisfy rapidity evolution equations governed by the Collins-Soper kernel $\Ktmd$. 
This is a striking result: despite the fully inclusive nature of DIS, its factorization in the threshold limit retains a structure characteristic of TMD dynamics. 
We will also show that, with an appropriate choice of the rapidity scale $y_n$, the functions $\mell{\bm{\mathcal{F}}}_{j}$ and $\mell{\bm{\mathcal{J}}}_j$ can be matched onto their light-cone counterparts, \textit{i.e.}, the standard PDFs $f_j$~\cite{Baulieu:1979mr,Curci:1980uw,Collins:1981uw,Collins:1989gx} and the inclusive jet functions $J_j$~\cite{Collins:2007ph,Accardi:2020iqn,Accardi:2023cmh} that also enter factorization formulas for event-shape observables~\cite{Bauer:2003pi,Schwartz:2007ib,vonManteuffel:2013vja,Bruser:2018rad}.
As a result, the factorized hadronic tensor in Eq.~\eqref{eq:fact_theo_sqrtdef} can be recast in the familiar light-cone form ~\cite{Chay:2012jr,Chay:2013zya,Chay:2017bmy,Fleming:2012kb}
\begin{align}
\label{eq:fact_theo_LC}
&W^{\mu\nu} = 
\mathbf{u}^{\mu \nu} \, H(\mu,Q) 
\invM 
\mell{f}_{j} (N;\mu)
    \,
\mell{J}_j(N;\mu).
\end{align}
From this perspective, threshold factorization in terms of light-cone operators emerges from a careful disentangling of collinear and soft contributions, where rapidity divergences and soft-collinear overlap subtractions are handled in close analogy with TMD factorization.

Although the inclusive factorization of Eq. \eqref{eq:fact_theo_LC} does not provide direct access to the Collins-Soper kernel, this remains encoded in the off-light-cone operator definitions of $\mell{\bm{\mathcal{F}}}_{j}$ and $\mell{\bm{\mathcal{J}}}_j$ and their rapidity evolution. 
This suggests that $\Ktmd$ might be accessible using collinear correlators in lattice QCD along the lines of existing quasi-TMD approaches~\cite{Bollweg:2024zet,Avkhadiev:2024mgd,Avkhadiev:2023poz,Shu:2023cot,Schlemmer:2021aij,Shanahan:2021tst,LatticeParton:2020uhz,LatticePartonLPC:2022eev,LatticeParton:2023xdl, Shanahan_2020,Ebert_2019,Ebert_2020}.

\section{Off-light-cone Factorization}
Similarly to SIDIS processes when the transverse momentum of the outgoing hadron with respect to the lepton plane is small compared to the hard scale $Q$, the final state in DIS at threshold develops near the plus (target) and minus (struck-quark jet) light-cone axes. 
Factorization can then be derived following the standard TMD procedure of Ref.~\cite{Collins:2011zzd}, as we briefly outline here.
\begin{figure}[tbhp]
\centering
\includegraphics[width=.35\textwidth]{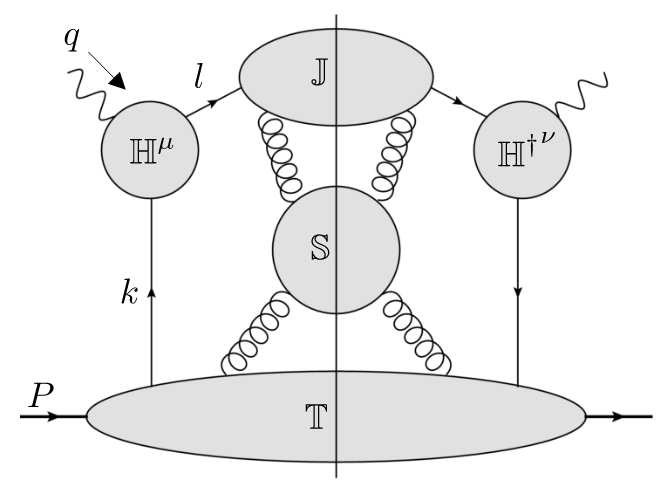}
\caption{DIS leading power region diagram at threshold. 
}
\label{fig:momentum_regions}
\end{figure}

The leading-power momentum regions are shown in Fig.~\ref{fig:momentum_regions}. Hard, far off-shell partons are absorbed into the photon--quark vertex subgraphs $\mathbb{H}$. The soft radiation is included in $\mathbb{S}$ and the collinear radiation in the target and jet subgraphs, $\mathbb{T}$ and $\mathbb{J}$, respectively. 
The kinematic approximations required to disentangle these regions involve lightlike projections that generate rapidity divergences at intermediate steps. 
Their consistent treatment requires the introduction of a rapidity regulator. 
In this work, we adopt an off-light-cone scheme~\cite{Ji:2004wu,Ji:2004xq,Chen:2006vd,Collins:2011zzd}, implemented by slightly tilting the lightlike directions associated with the approximated momenta. This choice ensures genuinely gauge-invariant operator definitions for the collinear and soft sectors, while preserving the underlying geometrical structure of the process.
In practice, we parametrize the tilted directions $n_1$ (collinear to the target) and $n_2$ (collinear to the jet) in terms of two rapidity regulators $y_1$ and $y_2$ as follows:
\begin{align}
    &n_1  =  (1,-e^{-2 y_1},\vec{0}_T) 
    \xrightarrow[y_1 \to +\infty]{}  (1,0,\vec{0}_T)\, ,
    \notag \\
    &n_2  = (e^{2 y_2},1,\vec{0}_T) 
    \quad \xrightarrow[y_2 \to -\infty]{}  (0,1,\vec{0}_T)\, .\
    \label{def:tilts}
\end{align}
The space-like tilt on the target side is chosen to resemble the virtual struck quark, while the time-like tilt on the jet side is chosen to resemble the outgoing final-state parton. 
Factorization then proceeds by approximating the gluons connecting hard, soft and collinear subgraphs as eikonal particles along the tilted directions, and collecting them into Wilson lines operators by means of QCD Ward identities.
The jet-side Wilson lines are future-pointing, as required to avoid contributions from the Glauber region in loop integrals~\cite{Collins:1997sr,Collins:2004nx,Collins:2011zzd}.
On the target side, we use past-pointing Wilson lines, as naturally implied by the eikonalization of the incoming parton~\footnote{This choice is a viable alternative to the corresponding future-pointing formulation~\cite{Collins:2004nx,Collins:2011zzd}, which is often preferred when discussing universality with final-state processes such as \texorpdfstring{$e^+e^-$}{e+e-} annihilation.}. 
After subtracting the double counting from soft radiation overlapping with the collinear sectors, we find that the hadronic tensor factorizes into a hard function $H_j$ and a double convolution involving a soft function $\mathcal{S}$, a target function $\mathcal{F}_{i}$, and a jet function $\mathcal{J}_i$ (that are not yet the boldface functions appearing in Eq.~\eqref{eq:fact_theo_sqrtdef}):
\begin{align}
\label{eq:fact_theo_xspace}
    x \, W^{\mu\nu} & = 
        \mathbf{u}^{\mu \nu} \, H_j(Q^2) \, 
        \int_{x}^{1} \frac{d \xi}{\xi} \, \int_0^{\xi-x} d \rho\, \\
    & \times 
        \mathcal{F}_{j}(\xi; y_1) \,
        \mathcal{S}\Big(\frac{\rho}{\xi}; y_1, y_2\Big) \, 
        \mathcal{J}_j\Big( \frac{\xi-x-\rho}{x}; y_2 \Big),
        \notag
\end{align}
where the renormalization scale $\mu$ is suppressed for simplicity. 
Here, $\xi$ denotes the momentum fraction of the target quark with respect to the incoming hadron, while $\rho$ parametrizes the fraction carried by soft radiation relative to the target quark. 
The factorization formula~\eqref{eq:fact_theo_xspace} exhibits a structure reminiscent of the classic  threshold factorization derived by Sterman~\cite{Sterman:1986aj}, with the key difference that the soft and collinear functions are defined \textit{off the light cone}, as reflected in their dependence on the rapidity regulators. 
Their operator definitions are given below, along with the needed soft-collinear subtractions.

\begin{figure}[H]
\centering
\includegraphics[width=.35\textwidth]{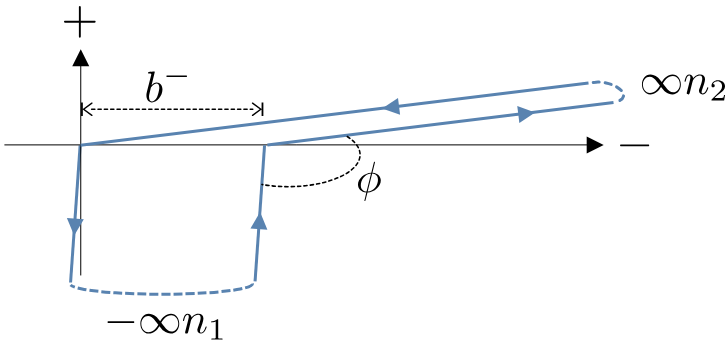}
\caption{
Wilson loop associated with the threshold DIS soft function. The displacement $b^-$ and the cusp angle $\phi$ are indicated. The off-light-cone directions $n_1$ and $n_2$ are defined in Eq.~\eqref{def:tilts}. 
\label{fig:soft_DIS_loop}}
\end{figure}

\smallskip 
\noindent\textbf{Soft function }~---~The soft function $\mathcal{S}$ entering Eq.~\eqref{eq:fact_theo_xspace} is the vacuum expectation value of the Wilson loop along the path in Fig.~\ref{fig:soft_DIS_loop}, traced and averaged over color indices:
\begin{align}
    \label{eq:DIS_soft_function}
    &\mathcal{S}\Big(\omega; \mu, y_1, y_2\Big) = 
    Z_S\big(\eps; \alpha_s, \phi \big) 
    \int \frac{d N}{2\pi i} (1-\omega)^{-N} \,
    \notag \\
    &\times
    (\text{Tr}/{N_c}) \,
    \langle 0|\mathcal{P}\text{exp}\left\{
    -i g_s^{(0)} \oint d x^\mu A_\mu^{(0)}(x)
    \right\} | 0 \rangle.
\end{align}
The Wilson lines are in the fundamental representation of $SU(N_c)$, and they are past-pointing along the target direction $n_1$ and future-pointing along the jet direction $n_2$.
Path ordering is indicated by the symbol $\mathcal{P}$.
The path depends on the Mellin variable $N$ through the relation $(b^- P^+)^2 = -N^2$, with $P$ the target momentum. In the threshold limit of large $N$, this corresponds to a large light-cone separation $b^-$~\footnote{In the threshold limit, the real part of \texorpdfstring{$N$}{N} can be neglected, making the displacement $b^-$ real.}. 
The superscript ``$(0)$'' indicates bare quantities.
The ultraviolet renormalization factor $Z_S$ depends only on the strong coupling, the cusp angle $\phi = \sinh^{-1}{n_1 \cdot n_2/\sqrt{-n_1^2\, n_2^2}}$, which becomes large and positive in the light-cone limit, and the dimensional 
regularization parameter $\eps$.
The geometry of the loop suggests that rapidity divergences are primarily associated with large positive rapidities, while large negative rapidities do not appear to generate additional singularities.
This feature can be traced back to the kinematics of Eq.~\eqref{eq:fact_theo_xspace}, where the convolution variables are directly associated with plus-momentum components in the Breit frame. In coordinate space, this implies that motion along the plus direction remains unconstrained, whereas motion along the minus direction is kinematically restricted.
This apparent irrelevance of large negative rapidities is consistent with the limit $y_2 \to -\infty$, which reproduces the soft factor of Korchemsky and Marchesini~\cite{Korchemsky:1992xv}, up to the different direction of the target-side tilt. 
Nevertheless, we retain an explicit regulator to preserve the geometrical structure of the problem and  make the connection with TMD factorization transparent.

In Mellin space, the evolution of the soft function with respect to the rapidity regulators,
\begin{align}
\label{eq:evo_soft_dis}
&\frac{\partial \log{\mell{\mathcal{S}(N,\mu,y_1,y_2)}}}{\partial y_{1(2)}}
= \pm \Kdis(\alpha_S, \mu\, l_N e^{y_{1(2)}}) + \dots
\end{align}
is governed by the kernel $\Kdis$, with $l_N = {N e^{\gamma_E}}/{\sqrt{2}\,P^+}$ denoting the characteristic length scale of DIS at threshold. 
The terms represented by the ellipses are required to preserve gauge invariance and are subleading in the light-cone limit of large rapidity separation.
In the next Section, we show that $\Kdis$ is not a new function, but in fact coincides with the Collins-Soper kernel upon replacing the characteristic transverse distance of TMD factorization with an appropriate combination of rapidity and characteristic longitudinal length of threshold DIS.

\smallskip 
\noindent\textbf{Target function }~---~Before soft-collinear subtraction, the unsubtracted $\mathcal{F}_{j}^{\text{\footnotesize \,uns}}$ target function is an off-light-cone generalization of the standard PDF:
\begin{align}
    \label{eq:offLC_TargetFun_uns}
    \mathcal{F}_{j}^{\text{\footnotesize \,uns}}(\xi;\mu,y_2) = 
    Z_T(\eps;\xi,\alpha_s,y_2)
    \int \frac{d \sigma^-}{2\pi} e^{-i \xi P^+ \sigma^-}
    \notag \\
    \times \mathrm{Tr}\,\langle P | 
    \overline{\psi}_j^{(0)}(\sigma)
    [\infty n_2 \to \sigma]
    \frac{\gamma^+}{2}
    [0 \to \infty n_2]
    \psi_j^{(0)}(0)
    | P \rangle ,
\end{align}
where $\sigma = (0,\sigma^-, \vec{0}_T)$ and the trace is over color indices. The gauge links are defined along the off-light-cone future-pointing direction $n_2$ and denoted by square brackets.
The factor $Z_T$ renormalizes ultraviolet divergences.
In the threshold limit, soft radiation factorizes and must be subtracted to avoid double counting in the cross section. The subtracted target function is then defined as
\begin{align}
    \label{eq:Fj_def}
    &\mathcal{F}_j(\xi,\mu,y_1) = 
    \int \frac{d N}{2\pi i} \xi^{-N} 
    \lim_{\eta \to -\infty} \frac{\mell{\mathcal{F}}_{j}^{\text{\footnotesize \,uns}} \left(N;\mu,\eta\right) }{\mell{\mathcal{S}(N,\mu,y_1,\eta)}} \,.
\end{align}
As a consequence of Eq.~\eqref{eq:evo_soft_dis}, the subtracted $\mathcal{F}_j$ target function evolves in rapidity with leading kernel $-\Kdis$.

\smallskip 
\noindent\textbf{Jet function }~---~The jet function is defined analogously. 
Before subtractions, it corresponds to an off-light-cone generalization of the invariant jet mass distribution:
\begin{align}
    \label{eq:offlc_JetFun_uns}
    &\mathcal{J}_j^{\text{\footnotesize \,uns}}(l^2/Q^2;\mu,y_1) = 
   Z_J(\eps; l^2/Q^2,\alpha_s,y_1)
   \frac{Q^2}{2\pi l^-} 
   \notag 
   \\
    &\quad\frac{\mathrm{Tr}}{N_c} \,
    \mathrm{Disc}
   \int \frac{d^4 \omega}{(2\pi)^4} e^{-i l \cdot \omega}
     \langle 0 |
     \mathcal{T} 
     \big\{
    [\omega \to -\infty n_1]
    \psi_j^{(0)}(\omega)\big\}
    \notag
    \\
    &\quad\times 
    \frac{\gamma^-}{4}
    \overline{\mathcal{T}}
    \big\{
    \overline{\psi}_j^{(0)}(0)
    [-\infty n_1 \to 0] 
    \big\}
    | 0 \rangle,
\end{align}
where $l^2$ denotes the jet invariant mass~\cite{Accardi:2008ne,Accardi:2020iqn,Accardi:2023cmh}, $Z_J$ renormalizes ultraviolet divergences, and gauge links are along the direction $n_1$. 
The symbols $\mathcal{T}$ and $\overline{\mathcal{T}}$ denote time- and anti-time-ordering.
Similarly to the target sector, subtracting the soft factor yields
\begin{align}
    \label{eq:Jj_def}
    &\mathcal{J}_j(z,\mu,y_2) = 
    \int \frac{d N}{2\pi i} \big(1+z\big)^{N} 
    \lim_{\eta \to +\infty} \frac{\mell{\mathcal{J}}_{j}^{\text{\footnotesize \,uns}} \left(N;\mu,\eta\right) }{\mell{\mathcal{S}(N,\mu,\eta,y_2)}} \,.
\end{align}
and the resulting jet function $\mathcal{J}_j$ evolves in rapidity with leading kernel $+\Kdis$.

\section{Emergence of the Collins-Soper kernel}
\begin{figure}[H]
\centering
\includegraphics[width=.35\textwidth]{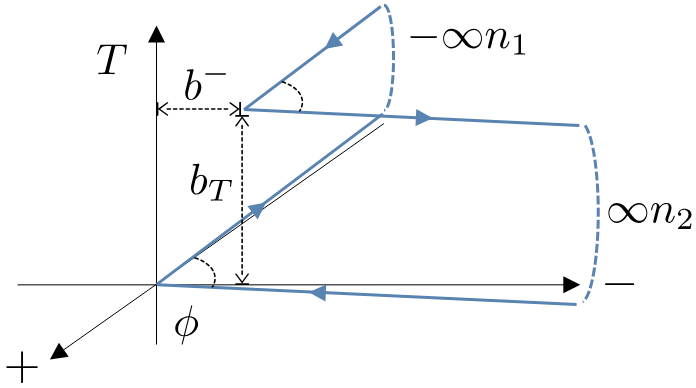}
\caption{
Wilson loop associated with the generalized soft function. The displacement $b$ has both a light-cone component $b^-$ and a transverse component $b_T$. The cusp angle $\phi$ is indicated.
\label{fig:generalized_soft}}
\end{figure}

We now show that the kernel $\Kdis$ introduced in the previous Section coincides with the Collins-Soper kernel $\Ktmd$ governing the rapidity evolution of TMDs. 
We give here the essential argument, while a detailed derivation will be presented in a forthcoming work.
To this end, building on similar strategies explored in the literature~\cite{Li:2016axz}, we introduce a generalized soft function $\mathcal{S}_{\text{\footnotesize gen}}$, defined as the vacuum expectation value of a Wilson loop characterized by a displacement $b$ with components both along the minus direction and the transverse plane, as illustrated in Fig.~\ref{fig:generalized_soft}.
\noindent
This operator reduces to the threshold DIS soft function in the limit $b_T \to 0$, and to the SIDIS TMD soft factor in the limit $b^- \to 0$. 
Exploiting standard results on the renormalization of Wilson lines~\cite{Polyakov:1980ca,Korchemsky:1987wg}, together with the constraints imposed by the choice of light-cone and transverse directions, we find that the rapidity evolution of the generalized soft function reads
\begin{align}
\label{eq:evo_soft_gen}
\frac{\partial \log{\mell{\mathcal{S}_{\text{\footnotesize gen}}(b,\mu,y_1,y_2)}}}{\partial y_{1(2)}}
= \pm \Kgen(\alpha_s(\mu),\mu, l_{1(2)}, l_T) + \dots,
\end{align}
where $l_{1(2)} = (b^- e^{\gamma_E}/\sqrt{2}) e^{y_{1(2)}}$ and $l_T = b_T e^{\gamma_E}/2$ denote the characteristic longitudinal and transverse lengths entering the problem.

The geometry of the Wilson loop, combined with the light-cone limit intrinsic to the definition of $\Kgen$, implies the renormalization group equation ${d\Kgen}/{d\log\mu}=-2\Gamma_K(\alpha_s)$, 
where $\Gamma_K$ is the universal light-like cusp anomalous dimension and the factor of two reflects the presence of the two cusps shown in Fig.~\ref{fig:generalized_soft}.
Thus, at each order in perturbation theory the renormalization scale $\mu$ enters $\Kgen$ through the coupling $\alpha_s$ and logarithms of $(\mu(l \pm i\,0))$, where $l$ denotes either of the two characteristic lengths, $l_{1(2)}$ or $l_T$, and the imaginary prescription is fixed by the pole structure of the gluon propagator in coordinate space~\cite{Korchemsky:1992xv}. 
The remaining length can only appear through dimensionless ratios entering the coefficients of the logarithms. 
At lowest-order (1-loop), we find
\begin{align}
    &\Kgen
    =
    -\frac{\alpha_s}{\pi} C_F
    \bigg[
    2 L 
    + 
    \label{eq:Kgen_1loop} \\
    &\,+
    \log{r^2} + \frac{1}{\sqrt{1-4 r^2}}\log{\Big(\frac{1-2r^2 + \sqrt{1-4r^2}}{2 r^2}\Big)}
    \bigg] 
    + \mathcal{O}(\alpha_s^2)
    \notag
\end{align}
where $L = \log{\big(\mu(l_{1(2)} - i\,0)\big)}$ and $r = (l_T + i\,0)/(l_{1(2)} - i\,0)$.
Equivalently, the result can be expressed in terms of $\log{\big(\mu(l_T + i\,0)\big)}$, with the dependence on $l_{1(2)}$ entering only through the inverse dimensionless ratio.
Indeed, $\Kgen$ admits two equivalent analytic representations, corresponding to distinct branches related by analytic continuation. 
This realizes a form of crossing symmetry between $l_{1(2)}$ and $l_T$ scales, analogous to the symmetry between the $s$- and $t$-channels in $2 \to 2$ scattering amplitudes.
Comparing now the limit $l_T \to 0$ in the two representations, one obtains the equivalence between the rapidity kernels of threshold DIS and TMD SIDIS to all orders:
\begin{align}
    \label{eq:equivalence}
    \Kdis(\alpha_s,\mu\,l_{1(2)}) = \Ktmd(\alpha_s,\mu\,l_{1(2)}) \, ,
\end{align}
which we have explicitly verified at one loop. 
Thus, the DIS rapidity kernel in Eq.~\eqref{eq:evo_soft_dis} is in fact the Collins-Soper kernel evaluated at the characteristic length $l_N e^{y_{1,2}}$ rather than the transverse separation $b_T$ of a TMD process.
This reveals the universality of the kernel, and shows that the soft radiation emitted between two widely separated light-cone directions in seemingly unrelated processes is governed by the same function.

\section{Rapidity evolution and matching to the light cone}
\label{sec:rapidity_evolution}

The role the Collins-Soper kernel plays in filling the rapidity gap with soft radiation becomes manifest upon a suitable redefinition of the target and jet functions:
\begin{strip}
    \begin{align}
    &\mell{\bm{\mathcal{F}}}_{j}(N;\mu,y_n) = 
    \lim_{\substack{\eta \to +\infty \\ \overline{\eta} \to -\infty}}
    \mell{\mathcal{F}}_{j}^{\text{\footnotesize \,uns}} \left(N;\mu,\overline{\eta}\right)
    \sqrt{\frac{\mell{\mathcal{S}(N,\mu,\eta,y_n)}}{\mell{\mathcal{S}(N,\mu,\eta,\overline{\eta})}
    \mell{\mathcal{S}(N,\mu,y_n,\overline{\eta})}}}\,,
    \label{eq:sqrt_F}
    \\
    &\mell{\bm{\mathcal{J}}}_{j}(N;\mu,y_n) = 
    \lim_{\substack{\eta \to +\infty \\ \overline{\eta} \to -\infty}}
    \mell{\mathcal{J}}_{j}^{\text{\footnotesize \,uns}} \left(N;\mu,\eta\right)
    \sqrt{\frac{\mell{\mathcal{S}(N,\mu,y_n,\overline{\eta})}}{\mell{\mathcal{S}(N,\mu,\eta,\overline{\eta})}
    \mell{\mathcal{S}(N,\mu,\eta,y_n)}}}\,.
    \label{eq:sqrt_J}
    \end{align}
\end{strip}
\!\!This rearrangement, familiar from TMD factorization~\cite{Aybat:2011zv,Collins:2011zzd,Echevarria:2011epo}, recasts the factorization of Eq.~\eqref{eq:fact_theo_xspace} into the simple form of Eq.~\eqref{eq:fact_theo_sqrtdef}, turning the double convolution into a single one. 
More importantly, the subleading contributions cancel, leaving a rapidity evolution entirely governed by the Collins-Soper kernel:
\begin{subequations}
\label{eq:sqrt_CSevo}
    \begin{align}
        &\frac{\partial \log{\mell{\bm{\mathcal{F}}}_{j}(N;\mu,y_n)}}{\partial y_n}
        = -\Ktmd\big(\alpha_s, \mu l_N e^{y_n} \big)\,;
        \\
        &
        \frac{\partial \log{\mell{\bm{\mathcal{J}}}_{j}(N;\mu,y_n)}}{\partial y_n}
        = +\Ktmd\big(\alpha_s, \mu l_N e^{y_n} \big)\,.
    \end{align}
\end{subequations}
The rapidity evolution of the target and jet functions therefore contains an intrinsically nonperturbative component, as previously noted in the literature~\cite{Fleming:2012kb,Fleming:2016nhs}. 
Rather than signaling a breakdown of factorization or a violation of universality~\cite{Fleming:2012kb}, our analysis shows that this component is entirely encoded in a universal function governing the soft radiation, identified at operator level with the Collins-Soper kernel.
This perspective suggests that $\Ktmd$ may also play a role in other inclusive reactions characterized by rapidity gaps, such as threshold Drell-Yan production and dijet configurations in electron-positron collisions.

Equations~\eqref{eq:sqrt_CSevo} also suggest a new avenue for nonperturbative calculations of the Collins-Soper kernel using the newly defined off-light-cone collinear correlators~\eqref{eq:sqrt_F} and~\eqref{eq:sqrt_J}.
Since these correlators closely resemble the operator structures appearing in TMD factorization, it is natural to ask whether they admit a Euclidean formulation suitable for lattice calculations. If so, one could envision extracting the Collins-Soper kernel from the dependence of the corresponding quasi-collinear matrix elements on the rapidity tilts, along lines similar to existing quasi-TMD approaches~\cite{Bollweg:2024zet,Avkhadiev:2024mgd,Avkhadiev:2023poz,Shu:2023cot,Schlemmer:2021aij,Shanahan:2021tst,LatticeParton:2020uhz,LatticePartonLPC:2022eev,LatticeParton:2023xdl, Shanahan_2020,Ebert_2019,Ebert_2020}. 
We leave a detailed investigation of this possibility to future work.

In order to solve the evolution equations~\eqref{eq:sqrt_CSevo} one needs to specify initial rapidity scales for the target and jet functions.
Near threshold, the characteristic length $l_N$ is large and positive, which suggests to start the rapidity evolution from $y_n = + \log{(\mu l_N)}$ for the target function and from $y_n = -\log{(\mu l_N)}$
for the jet function.
\begin{figure*}[tbh]
\centering
\includegraphics[width=0.49\textwidth]{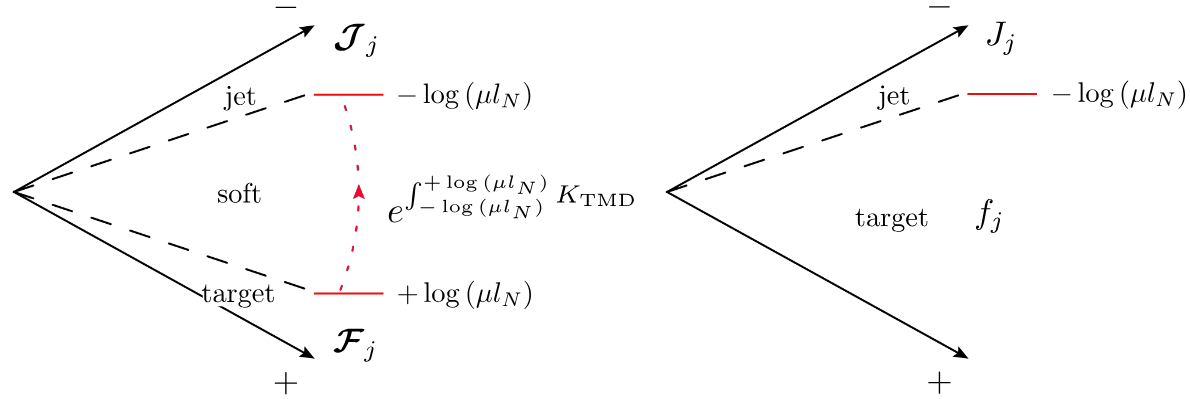}
\caption{
Pictorial representation of the rapidity scale choices discussed in the text. }
\label{fig:rapgraph}
\end{figure*}
With this choice, depicted in the left panel of Fig.~\ref{fig:rapgraph}, the role of the Collins-Soper kernel as the operator bridging the rapidity gap becomes explicit:
\begin{strip}
   \begin{align}
\label{eq:fact_theo_Nspace}
&W^{\mu\nu} = 
\mathbf{u}^{\mu \nu}\, H(\mu,Q) \, 
\invM
\mell{\bm{\mathcal{F}}}_{j}(N;\mu,+\log{(\mu l_N)})\, \mathrm{exp}\Big\{
    \int_{-\log{(\mu l_N)}}^{+\log{(\mu l_N)}} d y \, \Ktmd\big(\alpha_s, \mu \, l_N e^y\big)
    \Big\}
\mell{\bm{\mathcal{J}}}_{j}(N;\mu,-\log{(\mu l_N)})\,.
\end{align} 
\end{strip}
This expression parallels the TMD SIDIS factorization theorem, where the Mellin transform is replaced by a Fourier transform and the characteristic length becomes ${l_b = b_T/(2 e^{-\gamma_E})}$. In that case, one recovers the familiar factor $\exp[2 \log{(\mu l_b)} \, \Ktmd(\alpha_s, \mu b_T)]$ governing the $Q$ modulation of the low-$q_T$ cross section, while the target and jet functions $\bm{\mathcal{F}}$ and $\bm{\mathcal{J}}$ map onto the TMD parton distribution and fragmentation functions, respectively.

The rapidity scale $y_n = -\log{(\mu l_N)}$, adopted as a reference for the jet function, is peculiar. From inspecting the right-hand side of Eqs.~\eqref{eq:sqrt_CSevo}, it coincides with the value that removes all logarithms order by order in perturbation theory, reducing the Collins-Soper kernel to the rapidity anomalous dimension (RAD) $\gamma_r$~\cite{Li:2016ctv}, i.e. a simple perturbative series in $\alpha_s$. 
Remarkably, this holds both for the jet side and for the target side. 
In fact, at this rapidity scale, the square-root factors in both Eqs.~\eqref{eq:sqrt_F} and~\eqref{eq:sqrt_J} simplify significantly, leading to the matching relations
\begin{align}
\mell{\bm{\mathcal{F}}}_{j}(N;\mu,-\log{(\mu l_N)}) &= \frac{\mell{f}_{j}(N;\mu)}{C(\alpha_s, \mu \, l_N)} \,,
\\
\mell{\bm{\mathcal{J}}}_{j}(N;\mu,-\log{(\mu l_N)}) & = C(\alpha_s, \mu \, l_N) \,\mell{J}_j(N;\mu) \,.
\end{align}
Here, $\mell{f}_{j}(N;\mu)$ are the standard PDFs and $\mell{J}_j(N;\mu)$ the jet's invariant mass distribution. 
These relations explicitly match the off-light-cone functions $\bm{\mathcal{F}}_{j}$ and $\bm{\mathcal{J}}_{j}$ to their light-cone counterparts.
The matching coefficient $C$ has an anomalous dimension given by $-\tfrac{1}{2}\gamma_s$, where $\gamma_s$ is the universal soft anomalous dimension (SAD), that starts at two loops and corresponds to the RAD in conformal theories~\cite{Vladimirov:2016dll,Vladimirov:2017ksc}. 
At one loop, we have found explicitly ${C = 1 + \alpha_s C_F\, \pi/24}$. 

In light of this discussion, choosing the initial rapidities equal to $-\log{(\mu l_N)}$ for both the target and the jet function, as depicted in the right panel of Fig.~\ref{fig:rapgraph}, reproduces the factorization theorem~\eqref{eq:fact_theo_LC}, expressed entirely in terms of light-cone operators. 
In less inclusive cases, such as TMD factorization, a rapidity scale that simultaneously removes logarithmic structures in both the plus and minus directions cannot, in general, be chosen.

\section{Conclusions}
We have developed a novel off-light-cone factorization procedure for inclusive DIS at threshold that explicitly tracks soft gluon interactions across the rapidity gap between the target and final state jets. Exploiting geometrical similarities with TMD processes, we have provided a factorization theorem written in terms of newly defined collinear, \textit{off-light-cone} target and inclusive jet operators. This clarifies the validity of formulas previously derived on the light cone in QCD and SCET approaches, which emerge as special cases of our off-light-cone theorem. 
In particular, we have verified that our results are consistent with existing perturbative analyses, including those of Korchemsky and Marchesini~\cite{Korchemsky:1992xv} and of Fleming and Labun~\cite{Fleming:2012kb,Fleming:2016nhs} which can be interpreted as specific realizations of our framework.
Remarkably, the rapidity evolution of the off-light-cone operators is governed by the Collins-Soper kernel, revealing its role as a universal object controlling soft radiation. 
This also suggests a possible avenue for its determination in lattice QCD through a suitable Euclidean formulation of these correlators, to be explored in future work.

\vfill

\section*{Acknowledgements}
We gratefully acknowledge discussions with Johannes Michel and Ted Rogers. We also thank our theory group colleagues at Jefferson Lab for their feedback. This material is based upon work supported by the U.S. Department of Energy, Office of Science, Office of Nuclear Physics under Contract No. 89243126CSC000213 and No. DE-SC0025004, and by the European Union's ``Next Generation EU'' program through the Italian Ministry of University and Research (MUR) PRIN 2022 grants No. 2022SNA23K and No. 20225ZHA7W. 

\appendix
\section{Gauge invariance and subleading terms}
\label{app:gi}

\begin{figure}[H]
\centering
\includegraphics[width=0.15\textwidth]{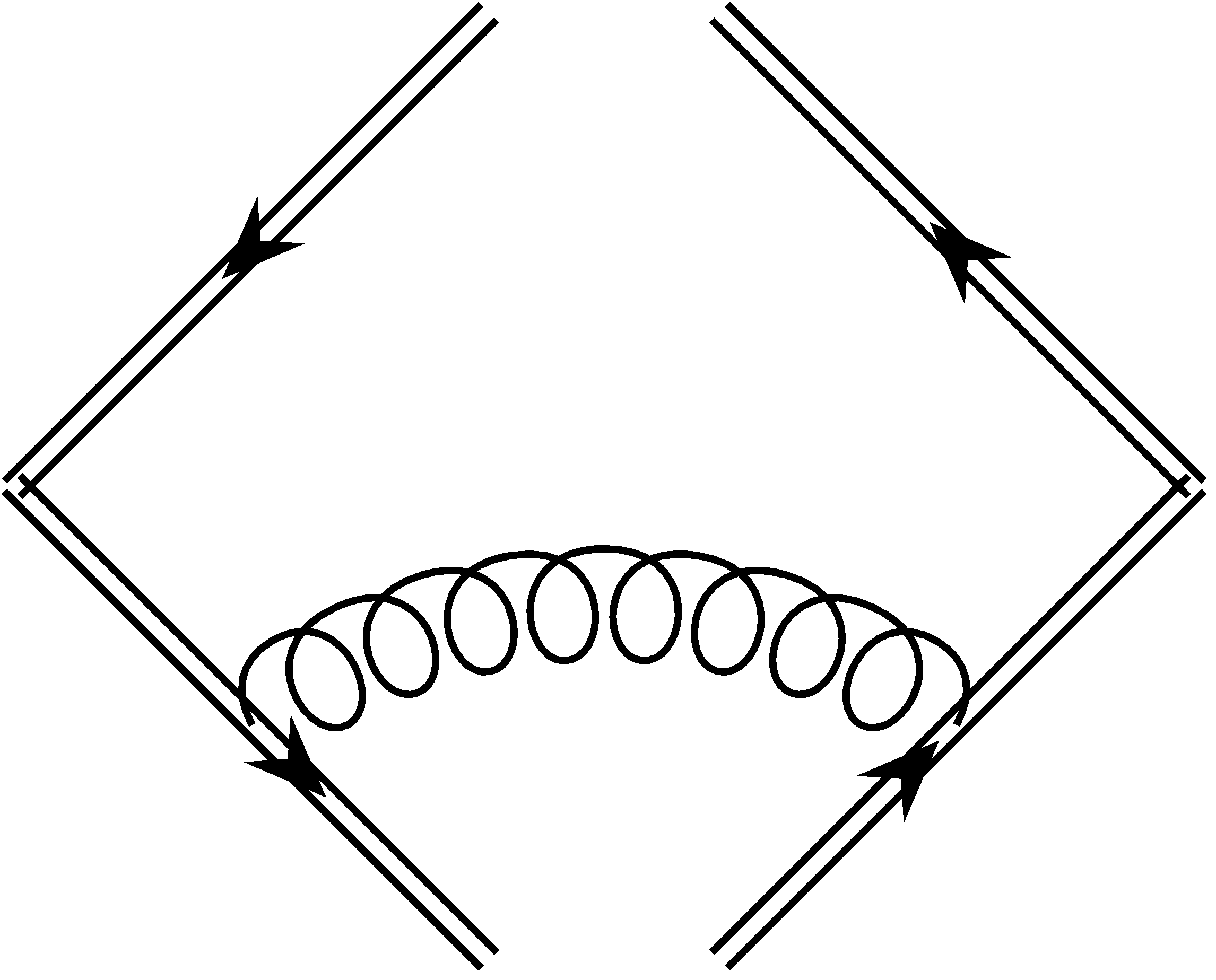}
\hspace{.5cm}
\includegraphics[width=0.15\textwidth]{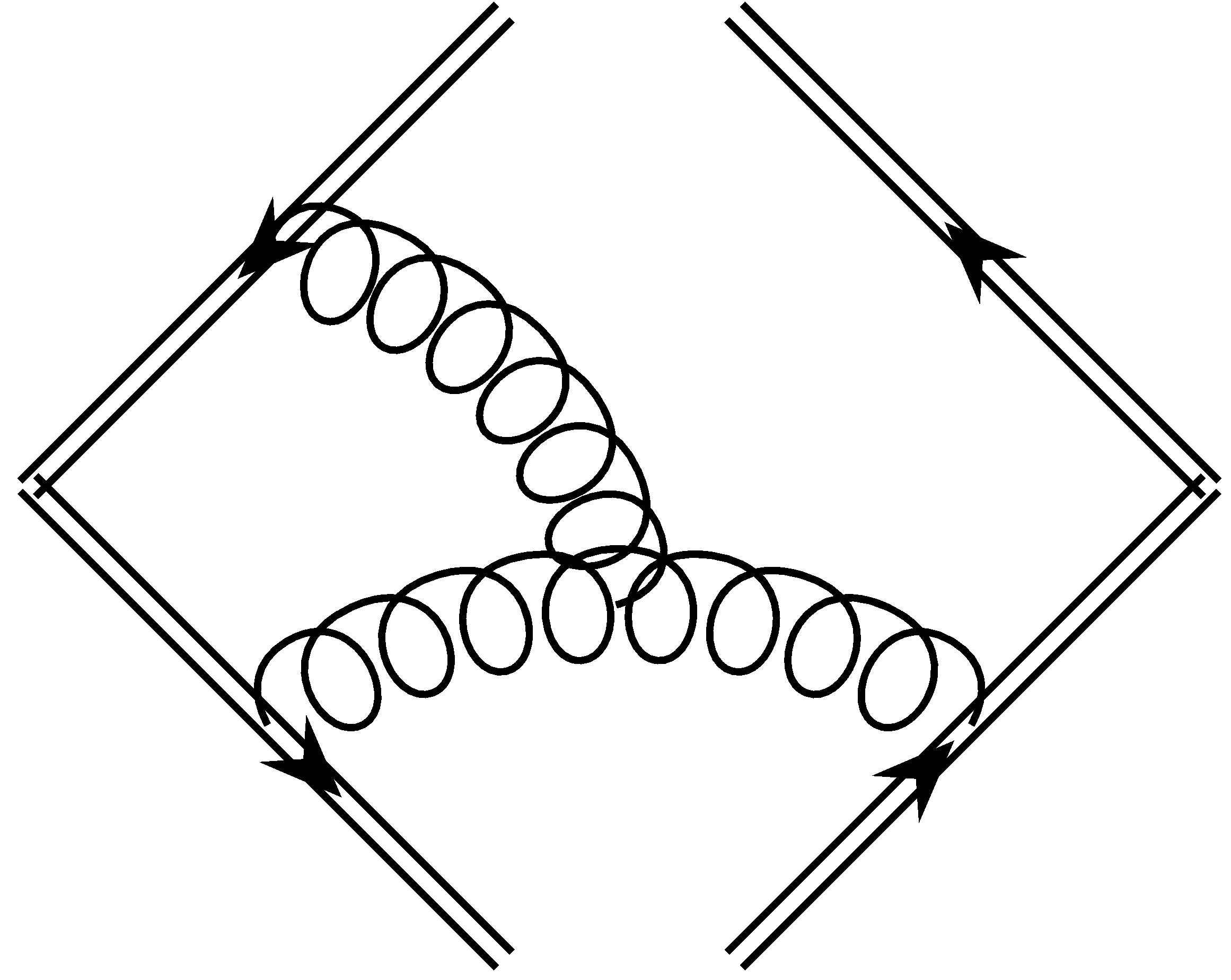}
\caption{
Representative diagrams contributing to subleading terms in the soft function in the light-cone limit.
}
\label{fig:subleading}
\end{figure}

In this Appendix, we discuss the gauge invariance of the operator definitions entering our factorization theorem, and the relation between gauge invariance and subleading terms in the light-cone limit, such as the ellipses in Eq.~\eqref{eq:evo_soft_dis}. 
In evaluating the soft function, the diagrams shown in Fig.~\ref{fig:subleading} contain contributions arising from soft gluon exchange between Wilson lines directed along the same direction $n^\mu$. 
Such terms are typically neglected in the literature, as they are proportional to $n^2$ and are therefore expected to vanish in the strict light-cone limit.
In fact, these contributions cancel in physical observables, but they are essential to ensure genuine gauge invariance of the operator definitions entering the factorization theorem. 
In particular, in non-covariant gauges the separation between leading and subleading contributions is no longer manifest, and diagrams of the type shown in Fig.~\ref{fig:subleading} give rise to leading rapidity divergences. 
In such cases, it becomes necessary to include the full set of contributions.
In this regard, we have verified, in direct comparison with the classic analysis of Ref.~\cite{Sterman:1986aj}, that only the inclusion of all such terms restores gauge invariance at the operator level.
At the same time, these contributions are ill-defined on the light cone: the factor $n^2$ multiplies a rapidity-divergent integral, leading to terms of the form $0\times\infty$. 
Thus, explicit calculations using light-cone rapidity regulators, such as the $\delta$-regulator~\cite{Echevarria:2011epo,Echevarria:2012js,Echevarria:2015usa,Echevarria:2015byo,Echevarria:2016scs} or the rapidity renormalization group~\cite{Chiu:2011qc,Chiu:2012ir}, might not assign a consistent value to these terms.
In our framework, the factorization theorem is formulated off the light cone, allowing these contributions to be defined in a natural and unambiguous way.
More specifically, the threshold DIS soft function introduced in Eq.~\eqref{eq:DIS_soft_function}, in the light-cone limit, reads in Mellin space:
\begin{align}
    \label{eq:soft_dis_withP}
    &\mell{\mathcal{S}}(N,\mu,y_1,y_2) = 
    \exp\big\{ 
    \int_{y_2}^{y_1} dy \, \Ktmd(\alpha_s, \mu l_N e^y)
    \notag \\
    &\quad+
    \frac{1}{2} \big[ 
    P(\alpha_s, \mu l_N e^{y_1})
    +
    P(\alpha_s, \mu l_N e^{y_2})
    \big]
    \big\},
\end{align}
where the subleading contributions are encoded in the function $P$. 
While the leading term, governed by the Collins-Soper kernel, is extensive in the rapidity interval, the subleading term appears instead as a boundary contribution. The same mechanism relating the leading kernels of threshold DIS and SIDIS TMD extends to the $P$-term, indicating that both bulk and boundary contributions are controlled by the same underlying structure.
An analogous decomposition holds for the target and jet functions of Eqs.~\eqref{eq:Fj_def} and~\eqref{eq:Jj_def}, which also contain corresponding $P$-terms. 
However, in the redefined operator combinations involving square-root soft factors of Eqs.~\eqref{eq:sqrt_F} and~\eqref{eq:sqrt_J}, these contributions cancel exactly, leaving a rapidity evolution governed solely by the Collins-Soper kernel. 
Should a suitable Euclidean formulation of these correlators be established, this cancellation could allow the Collins-Soper kernel to be isolated from ratios of matrix elements calculated in lattice QCD.

\bibliographystyle{elsarticle-num}
\biboptions{sort&compress}
\bibliography{bibliography}

\end{document}